\documentclass[%
 reprint,
 amsmath,amssymb,
 aps,
prstper,
]{revtex4-2}

\usepackage{graphicx}% Include figure files
\usepackage{dcolumn}% Align table columns on decimal point
\usepackage{bm}% bold math
\begin{document}

\preprint{AddendumAmJPhysConceptsFirstPaper}

\title{Addendum to Concepts First Paper:\\A Student Deficit Model is Untenable in Understanding a Demographic Grade Gap}% Force line breaks with \\

\author{David J. Webb}
\affiliation{%
 Dept. of Physics, Univ. of California, Davis, Davis CA
}%

\date{\today}% It is always \today, today,

\begin{abstract}
This addendum shows that a demographic grade gap is best understood using a course deficit model rather than the more common student deficit model.  Students from a concepts-first class took the same final exam as the three other classes that were offered at the same time so the grade gaps between different demographic groups from each class can be compared.  It is found that students identifying as members of a racial/ethnic group underrepresented in physics had higher final exam grades than their peers from the same concepts first class rather than the usual case, lower exam grades than their peers, found in each of the other three classes.  After controlling for demonstrated incoming understanding of calculus and Newtonian mechanics this demographic gap in the concepts first class is found to differ from that of the other three classes by over four standard errors.  In addition, it's noted that while students' incoming understandings are useful in explaining within-group differences they do not explain between-group differences suggesting that a student deficit model does not explain (between-group) demographic grade gaps.
\end{abstract}

\maketitle

%\tableofcontents

\section{\label{sec:Intro}Introduction}

Reference \cite{Webb2017} described an experimental class and a control class taught by the same instructor, covering the same topics, and even using exactly the same curricular materials.  The main difference between these two introductory Newtonian mechanics classes lay in the order of the presentation of material.  In the control class the topics of the course were covered chapter by chapter as usual.  In this chapter-by-chapter style the conceptual issues are studied together with simple calculations and then quickly followed by complicated calculations.  On the other hand, the experimental class spent the first 60\% of the course in an intensive study of all of the conceptual issues (along with very simple calculations) in the course.  The final 40\% of the course was then spent on the complicated calculations.  At the same time that these two classes were given there were also two other classes covering the same material but with two different instructors.

Reference \cite{Webb2017} compared final exam results for the control and experimental classes.  In order to avoid issues\cite{Gutierrez2008} associated with between-demographic-group comparisons (also known as gap gazing\cite{Rodriguez2001}) the paper mainly showed within-demographic-group final exam grade differences between the two classes.  It was found that each  demographic group studied had higher or equal final exam grades when they were enrolled in the concepts-first class than when they were enrolled in the control class.  In addition to these detailed within-group results we noted that the course grades of concepts-first students from racial/ethnic groups historically underrepresented in physics were indistinguishable from their peers in the same concepts-first class but that the usual result (lower than their peers) obtained in each of the other three classes.  This between-group result was noted but not carefully quantified.  This addendum is aimed at better quantifying this between-group affect.

The main reason for examining between-group affects is to add a voice to the growing body of recent research \cite{Salehi2019}\cite{Simmons2020}\cite{Shafer2021} on between-group demographic studies.  Specifically, some recent studies seem to take a Student Deficit Model\cite{Valencia1997} of demographic grade gaps and some studies seem to take a Course Deficit Model\cite{Cotner2017} of demographic grade gaps.  The work described in reference \cite{Webb2017} seems likely to support the Course Deficit Model.

\section{\label{sec:DataAndMethod}Data Set and Method}

The three other classes offered at the same time as the concepts-first class took the same final exam at the same time and they were all graded at the same time.  Unfortunately, the final exam had problems on it that the experimental and control groups had not seen but that one or both of the other classes had seen.  Because this makes a direct comparison of all four classes difficult and the fact that the comparison between the experimental and control classes told a fairly complete story, we left aside a more complete analysis using the other two classes.  In this addendum we will use Hierarchical Linear Modeling (HLM) to bring all of the classes into a single model that allows us to use all of our data in a comparison a particular demographic grade gap across two ways of organizing the course materials in an introductory physics course.

Using HLM allows one to take account of the fact that students are aggregated into classes so that the demographic grade gap calculation is first done at the class level using student-level variables before assembling the results from the different classes and including any class-level variables.  Fitting the data in this order helps one account for class-to-class variations (i.e. some classes have already seen some of the final exam questions).  In using HLM the student variables (FCI and math scores, and ethnicity) are at the lowest level and the class variable, concepts first or not, is at a higher level in the hierarchy.

The anonymized database from the previous study had the students' Force Concept Inventory\cite{Hestenes1992Force} pre-scores ($FCIpre$), their average calculus grades normalized to standard deviation = 1 ($IntroMath$), their final exam scores normalized to standard deviation = 1 ($FinalExam$), and the student ethnicities for each of the four classes.  This gives us a total of 633 students in the database that we use in this addendum (152 in the concepts-first class, 160 in the control class, 163 in the third class, and 158 in the fourth class).  The 14\% of the students who were from underrepresented racial/ethnic groups are denoted with a variable $URM = 1$ and their peers have $URM = 0$.

\section{\label{sec:Results}Results}

First we can look at the demographic gaps for all four classes.  Table \ref{Tbl1} shows, for each of the four classes, the average difference between students who are members of racial/ethnic groups underrepresented in physics and their peers in the same class.  The average demographic gap for the concepts-first class does not fall within any of the 95\% confidence intervals of the gaps from the other three classes.  On the other hand, \textbf{each} of the other three classes have gaps which fall within the 95\% confidence intervals of \textbf{each} of those three classes and fall outside the 95\% confidence interval of the concept's first class.  For these reasons, in the sequel I will gather those three standard courses together into one group and compare those classes directly with the concept's first class to quantify the differences while controlling for the students' incoming understandings of calculus and physics.

\begin{table}[htbp]
\caption{The final exam grade gaps for each of the four classes offered as the same time.  The grades are normalized so that the standard deviation over all students taking the final exam is 1.  The gap is between students from racial/ethnic groups underrepresented in physics ($URM$) and their peers.  The gap is negative if URM students received lower final exam grades.  The concepts first class is class number 1 and the control class (same instructor) is class number 2 and the other two classes were given at the same time and took the same final exam.}
\label{Tbl1}
\begin{ruledtabular}
\begin{tabular}{c c c c}
\textbf{Class} & \textbf{Grade Gap} &\textbf{Error} & \textbf{95\% confidence interval}\\ 
\hline
1 & 0.18 & 0.25 & -0.32 $\rightarrow$ 0.67 \\
2 & -0.90 & 0.23 & -1.34 $\rightarrow$ -0.44 \\
3 & -0.70 & 0.18 & -1.06 $\rightarrow$ -0.34 \\
4 & -0.80 & 0.23 & -1.25 $\rightarrow$ -0.34 \\
\end{tabular}
\end{ruledtabular}
\end{table}

One way to use HLM to compare the course structures is to fit model shown in equation \ref{eqn:HLMModel1} separately for i) the concepts-first class and ii) the other three classes.  The resulting coefficients are shown in Table \ref{Tbl2}.

\begin{multline}
FinalExam = b_0 + b_{URM}(URM) \\ + b_{FCIpre}FCIpre + b_{IntroMath}IntroMath
\label{eqn:HLMModel1}
\end{multline}

\begin{table}[htbp]
\caption{The coefficients from HLM fitting of equation \ref{eqn:HLMModel1} for 1) the concepts-first class and 2,3,4) the other three classes fit as a group.}
\label{Tbl2}
\begin{ruledtabular}
\begin{tabular}{c c c c c}
\textbf{Coefficient} & \textbf{Class} & \textbf{Value} &\textbf{Error} & \textbf{95\% conf. intvl.}\\
\hline
$b_{URM}$ & 1 & 0.38 & 0.15 & 0.09 $\rightarrow$ 0.68 \\
 & 2,3,4 & -0.39 & 0.09 & -0.56 $\rightarrow$ -0.21 \\
\hline
$b_{FCIpre}$ & 1 & 0.089 & 0.007 & 0.07 $\rightarrow$ 0.10 \\
 & 2,3,4 & 0.062 & 0.005 & 0.05 $\rightarrow$ 0.07 \\
\hline
$b_{IntroMath}$ & 1 & 0.56 & 0.07 & 0.41 $\rightarrow$ 0.70 \\
 & 2,3,4 & 0.62 & 0.05 & 0.52 $\rightarrow$ 0.71 \\
\hline
$b_{0}$ & 1 & -1.59 & 0.13 & -1.84 $\rightarrow$ -1.34 \\
 & 2,3,4 & -1.09 & 0.12 & -1.34 $\rightarrow$ -0.85 \\
\end{tabular}
\end{ruledtabular}
\end{table}

The important coefficient, $b_{URM}$, measures the demographic final exam grade gap.  As we noticed from the uncorrected gap, the gaps in the two different organizations of class material are of opposite sign.  Calculating the difference in the two values of the gap we find that the difference is 0.77 $\pm$ 0.17.  For this comparison the t-statistic = 4.43 so the two values of demographic gap differ by 4.43 standard errors.  This difference is almost as large as the difference (5 standard errors) that physicists use to identify a new ``discovery''.  At this point we should note that the student-level variables $FCI$ and $IntroMath$ are excellent predictors of differences between the students in a particular demographic group and so are good within-group predictors.  Nevertheless, correcting for these student-level variables not only does not explain (i.e. reduce to zero) the demographic gap but may have enlarged the demographic gap in the concepts-first class.  So using these particular student-level variables as between-group predictors is untenable for these data.

A possible issue with the above fitting is that the two types of class organization are treated separately, and so differently.  Interestingly, the $FCIpre$ coefficients were measurably different from each other using this method.  We can get around this possible problem by simply treating the class differences within a single model fitting the entire database.  To do this we define a second categorical variable, $Course$, which equals 0 for the group of classes taught in the usual way and equal to 1 for the concepts-first class and we include an interaction term, $URM*Course$ to measure a difference between the two class structures on the resulting demographic gap.  The variable $Course$ is a class-level variable and a further importance of HLM is that class-level variables are treated so that the error estimates automatically take into account the fact that this variable is correlated at the class-level.  We use HLM to fit equation \ref{eqn:HLMModel2}

\begin{multline}
FinalExam = b_0 \\ + b_{URM}(URM) + b_{Course}(Course) \\ + b_{URM*Course}(URM*Course)\\ + b_{FCIpre}FCIpre + b_{IntroMath}IntroMath
\label{eqn:HLMModel2}
\end{multline}

\begin{table}[htbp]
\caption{The coefficients from HLM fitting of equation \ref{eqn:HLMModel2}.  The URM coefficient will give the demographic gap for the three classes offered with the standard organization and the interaction coefficient will give the difference between the standard demographic gap and the demographic gap from the concepts-first class organization.}
\label{Tbl3}
\begin{ruledtabular}
\begin{tabular}{c c c c c}
\textbf{Coefficient} & \textbf{Value} &\textbf{Error} & \textbf{z-statistic} & \textbf{P-value}\\
\hline
$b_{URM}$ & -0.38 & 0.09 & -4.36 & $<10^{-3}$ \\
$b_{Course}$ & -0.10 & 0.17 & -0.55 & 0.585 \\
$b_{Course*URM}$ & 0.77 & 0.19 & 4.12 & $<10^{-3}$ \\
$b_{FCIpre}$ & 0.069 & 0.004 & 16.76 & $<10^{-3}$ \\
$b_{IntroMath}$ & 0.61 & 0.04 & 14.67 & $<10^{-3}$ \\
$b_{0}$ & -1.19 & 0.11 & 11.00 & $<10^{-3}$ \\
\end{tabular}
\end{ruledtabular}
\end{table}

Table \ref{Tbl2} tells the same story that we found above.  The demographic gap for the standard classes  is -0.38 and the gap in the concepts-first course is 0.77 higher than that of the standard classes.  So the two demographic gaps are still opposite in sign and with this calculation the t-statistic = 4.12 and comes out of the fit itself.  Again the t-statistic is fairly large and we find that using student-level variables as between-group predictors is untenable but that the differences may be explained by the course-level variable $Course$.

\section{\label{sec:Disc}Discussion}

In comparisons of physics grades, researchers often find different average grades for different demographic groups, these are between-group comparisons.  In addition, the size of these demographic grade gaps can often be reduced (sometimes reduced to essentially zero) by controlling for student-level variables that are related to academic skills, or academic preparation, or test-taking skills, etc. and that are known to help explain within-group grades.  Using student-level variables in this way can suggest that students from one demographic group may not have the same average skills or preparation as students from another demographic group.  This has been called a student deficit model of the demographic gap\cite{Valencia1997},\cite{Cotner2017}.

A different view of these demographic gaps comes when one considers that maybe all of the groups of students have the ability to learn and demonstrate that learning but that the course design and/or presentation may not be suited for equally teaching all of the different socioeconomic groups of students that make up the class.  This view suggests using a course deficit model\cite{Cotner2017} in explaining how it happens that different groups of students receive different grades.  Here, the word ``course'' is meant to encompass all of the details that are under the control of either the particular instructor or the school that offers the course.  Course differences could be as minor as changing the presentation of the material or changing the testing regime or include changing the order of the material presented, the number of classes a student takes, and/or the number of hours per class or include even more changes whose nature have not occurred to this author.

The question of deciding whether the student deficit model or the course deficit model is better at explaining why one demographic group outperforms another under a particular course regime is one that a growing group of researchers think should be considered.  The results of this study of the concepts-first class structure show that a particular demographics grade gap is not easily explained using a student deficit model that considers only preparation in physics (using the Force Concept Inventory) and preparation in math (using the students' introductory calculus scores).  Instead, the demographic grade gap seems highly correlated with the course structure so the data support a course deficit model of this gap.

\bibliography{addendum.bib}

\begin{thebibliography}{9}%
\makeatletter
\providecommand \@ifxundefined [1]{%
 \@ifx{#1\undefined}
}%
\providecommand \@ifnum [1]{%
 \ifnum #1\expandafter \@firstoftwo
 \else \expandafter \@secondoftwo
 \fi
}%
\providecommand \@ifx [1]{%
 \ifx #1\expandafter \@firstoftwo
 \else \expandafter \@secondoftwo
 \fi
}%
\providecommand \natexlab [1]{#1}%
\providecommand \enquote  [1]{``#1''}%
\providecommand \bibnamefont  [1]{#1}%
\providecommand \bibfnamefont [1]{#1}%
\providecommand \citenamefont [1]{#1}%
\providecommand \href@noop [0]{\@secondoftwo}%
\providecommand \href [0]{\begingroup \@sanitize@url \@href}%
\providecommand \@href[1]{\@@startlink{#1}\@@href}%
\providecommand \@@href[1]{\endgroup#1\@@endlink}%
\providecommand \@sanitize@url [0]{\catcode `\\12\catcode `\$12\catcode
  `\&12\catcode `\#12\catcode `\^12\catcode `\_12\catcode `\%12\relax}%
\providecommand \@@startlink[1]{}%
\providecommand \@@endlink[0]{}%
\providecommand \url  [0]{\begingroup\@sanitize@url \@url }%
\providecommand \@url [1]{\endgroup\@href {#1}{\urlprefix }}%
\providecommand \urlprefix  [0]{URL }%
\providecommand \Eprint [0]{\href }%
\providecommand \doibase [0]{https://doi.org/}%
\providecommand \selectlanguage [0]{\@gobble}%
\providecommand \bibinfo  [0]{\@secondoftwo}%
\providecommand \bibfield  [0]{\@secondoftwo}%
\providecommand \translation [1]{[#1]}%
\providecommand \BibitemOpen [0]{}%
\providecommand \bibitemStop [0]{}%
\providecommand \bibitemNoStop [0]{.\EOS\space}%
\providecommand \EOS [0]{\spacefactor3000\relax}%
\providecommand \BibitemShut  [1]{\csname bibitem#1\endcsname}%
\let\auto@bib@innerbib\@empty
%</preamble>
\bibitem [{\citenamefont {Webb}(2017)}]{Webb2017}%
  \BibitemOpen
  \bibfield  {author} {\bibinfo {author} {\bibfnamefont {D.~J.}\ \bibnamefont
  {Webb}},\ }\bibfield  {title} {\bibinfo {title} {{Concepts first: A course
  with improved educational outcomes and parity for underrepresented minority
  groups}},\ }\bibfield  {journal} {\bibinfo  {journal} {Am. J. Phys.}\
  }\textbf {\bibinfo {volume} {85}},\ \href {https://doi.org/10.1119/1.4991371}
  {10.1119/1.4991371} (\bibinfo {year} {2017})\BibitemShut {NoStop}%
\bibitem [{\citenamefont {Guti{\'{e}}rrez}(2008)}]{Gutierrez2008}%
  \BibitemOpen
  \bibfield  {author} {\bibinfo {author} {\bibfnamefont {R.}~\bibnamefont
  {Guti{\'{e}}rrez}},\ }\bibfield  {title} {\bibinfo {title} {A ``gap-gazing''
  fetish in mathematics education? problematizing research on the achievement
  gap},\ }\href {https://doi.org/10.2307/40539302} {\bibfield  {journal}
  {\bibinfo  {journal} {Journal for Research in Mathematics Education}\
  }\textbf {\bibinfo {volume} {39}},\ \bibinfo {pages} {357} (\bibinfo {year}
  {2008})}\BibitemShut {NoStop}%
\bibitem [{\citenamefont {Rodriguez}(2001)}]{Rodriguez2001}%
  \BibitemOpen
  \bibfield  {author} {\bibinfo {author} {\bibfnamefont {A.~J.}\ \bibnamefont
  {Rodriguez}},\ }\bibfield  {title} {\bibinfo {title} {{From gap gazing to
  promising cases: Moving toward equity in urban systemic reform}},\ }\href
  {https://doi.org/10.1002/tea.10005} {\bibfield  {journal} {\bibinfo
  {journal} {J. Res. Sci. Teach.}\ }\textbf {\bibinfo {volume} {38}},\ \bibinfo
  {pages} {1115} (\bibinfo {year} {2001})}\BibitemShut {NoStop}%
\bibitem [{\citenamefont {Salehi}\ \emph {et~al.}(2019)\citenamefont {Salehi},
  \citenamefont {Burkholder}, \citenamefont {Lepage}, \citenamefont {Pollock},\
  and\ \citenamefont {Wieman}}]{Salehi2019}%
  \BibitemOpen
  \bibfield  {author} {\bibinfo {author} {\bibfnamefont {S.}~\bibnamefont
  {Salehi}}, \bibinfo {author} {\bibfnamefont {E.}~\bibnamefont {Burkholder}},
  \bibinfo {author} {\bibfnamefont {G.~P.}\ \bibnamefont {Lepage}}, \bibinfo
  {author} {\bibfnamefont {S.}~\bibnamefont {Pollock}},\ and\ \bibinfo {author}
  {\bibfnamefont {C.}~\bibnamefont {Wieman}},\ }\bibfield  {title} {\bibinfo
  {title} {{Demographic gaps or preparation gaps?: The large impact of incoming
  preparation on performance of students in introductory physics}},\ }\href
  {https://doi.org/10.1103/physrevphyseducres.15.020114} {\bibfield  {journal}
  {\bibinfo  {journal} {Physical Review Physics Education Research}\ }\textbf
  {\bibinfo {volume} {15}},\ \bibinfo {pages} {20114} (\bibinfo {year}
  {2019})}\BibitemShut {NoStop}%
\bibitem [{\citenamefont {Simmons}\ and\ \citenamefont
  {Heckler}(2020)}]{Simmons2020}%
  \BibitemOpen
  \bibfield  {author} {\bibinfo {author} {\bibfnamefont {A.~B.}\ \bibnamefont
  {Simmons}}\ and\ \bibinfo {author} {\bibfnamefont {A.~F.}\ \bibnamefont
  {Heckler}},\ }\bibfield  {title} {\bibinfo {title} {{Grades, grade component
  weighting, and demographic disparities in introductory physics}},\ }\href
  {https://doi.org/10.1103/PhysRevPhysEducRes.16.020125} {\bibfield  {journal}
  {\bibinfo  {journal} {Physical Review Physics Education Research}\ }\textbf
  {\bibinfo {volume} {16}},\ \bibinfo {pages} {20125} (\bibinfo {year}
  {2020})}\BibitemShut {NoStop}%
\bibitem [{\citenamefont {Shafer}\ \emph {et~al.}(2021)\citenamefont {Shafer},
  \citenamefont {Mahmood},\ and\ \citenamefont {Stelzer}}]{Shafer2021}%
  \BibitemOpen
  \bibfield  {author} {\bibinfo {author} {\bibfnamefont {D.}~\bibnamefont
  {Shafer}}, \bibinfo {author} {\bibfnamefont {M.~S.}\ \bibnamefont
  {Mahmood}},\ and\ \bibinfo {author} {\bibfnamefont {T.}~\bibnamefont
  {Stelzer}},\ }\bibfield  {title} {\bibinfo {title} {{Impact of broad
  categorization on statistical results: How underrepresented minority
  designation can mask the struggles of both Asian American and African
  American students}},\ }\href
  {https://doi.org/10.1103/PhysRevPhysEducRes.17.010113} {\bibfield  {journal}
  {\bibinfo  {journal} {Physical Review Physics Education Research}\ }\textbf
  {\bibinfo {volume} {17}},\ \bibinfo {pages} {010113} (\bibinfo {year}
  {2021})}\BibitemShut {NoStop}%
\bibitem [{\citenamefont {Valencia}(1997)}]{Valencia1997}%
  \BibitemOpen
  \bibfield  {author} {\bibinfo {author} {\bibfnamefont {R.~R.}\ \bibnamefont
  {Valencia}},\ }\href@noop {} {\emph {\bibinfo {title} {{The Evolution of
  Deficit Thinking: Educational Thought and Practice}}}}\ (\bibinfo
  {publisher} {The Falmer Press},\ \bibinfo {address} {London},\ \bibinfo
  {year} {1997})\BibitemShut {NoStop}%
\bibitem [{\citenamefont {Cotner}\ and\ \citenamefont
  {Ballen}(2017)}]{Cotner2017}%
  \BibitemOpen
  \bibfield  {author} {\bibinfo {author} {\bibfnamefont {S.}~\bibnamefont
  {Cotner}}\ and\ \bibinfo {author} {\bibfnamefont {C.~J.}\ \bibnamefont
  {Ballen}},\ }\bibfield  {title} {\bibinfo {title} {{Can mixed assessment
  methods make biology classes more equitable?}},\ }\bibfield  {journal}
  {\bibinfo  {journal} {PLoS ONE}\ }\textbf {\bibinfo {volume} {12}},\ \href
  {https://doi.org/10.1371/journal.pone.0189610} {10.1371/journal.pone.0189610}
  (\bibinfo {year} {2017})\BibitemShut {NoStop}%
\bibitem [{\citenamefont {Hestenes}\ \emph {et~al.}(1992)\citenamefont
  {Hestenes}, \citenamefont {Wells},\ and\ \citenamefont
  {Swackhamer}}]{Hestenes1992Force}%
  \BibitemOpen
  \bibfield  {author} {\bibinfo {author} {\bibfnamefont {D.}~\bibnamefont
  {Hestenes}}, \bibinfo {author} {\bibfnamefont {M.}~\bibnamefont {Wells}},\
  and\ \bibinfo {author} {\bibfnamefont {G.}~\bibnamefont {Swackhamer}},\
  }\bibfield  {title} {\bibinfo {title} {{Force concept inventory}},\ }\href
  {https://doi.org/10.1119/1.2343497} {\bibfield  {journal} {\bibinfo
  {journal} {The Physics Teacher}\ }\textbf {\bibinfo {volume} {30}},\ \bibinfo
  {pages} {141} (\bibinfo {year} {1992})}\BibitemShut {NoStop}%
\end{thebibliography}%

\end{document}